\documentclass[pra,superscriptaddress,twocolumn,pra,longbibliography,floatfix,eqsecnum]{revtex4-2}
\usepackage{graphicx,amsmath,relsize,epstopdf,color,mathtools,bm,newtxtext,newtxmath,braket,rotating,dsfont}

\usepackage[colorlinks={true}, citecolor={blue}, filecolor={blue}, linkcolor={blue}, urlcolor={blue}]{hyperref}
\usepackage[caption=false]{subfig}

\usepackage{soul}
\usepackage{url}

\newcommand{\Tr}{\mathop{\mathrm{Tr}} \nolimits}

\allowdisplaybreaks
\DeclareMathOperator{\sech}{sech}

\begin{document}
\title{Coherence-Enhanced Spatial Quantum Thermometry}  
	\author{Asghar Ullah}
 \email{aullah21@ku.edu.tr}
\affiliation{Department of Physics, Ko\c{c} University, 34450 Sar\i yer, Istanbul, T\"urkiye}
%---------------------------------------------------%
\author{Giovanni Scala}
\affiliation{Dipartimento Interateneo di Fisica, Politecnico di Bari, 70126 Bari, Italy}
\affiliation{INFN, Sezione di Bari, 70126 Bari, Italy}
%\email{ giovanni.scala@poliba.it}
%---------------------------------------------------%
\author{Luis L. S\'{a}nchez-Soto}
\affiliation{Departamento de Óptica, Facultad de Física, Universidad Complutense, 28040 Madrid, Spain}
\affiliation{Max-Planck-Institut für die Physik des Lichts, 91058 Erlangen, Germany}
\affiliation{Institute for Quantum Studies, Chapman University, Orange, CA 92866, USA}

\author{\"Ozg\"ur E. M\"ustecapl\i o\u glu}	
	%\email{omustecap@ku.edu.tr}
	\affiliation{Department of Physics, Ko\c{c} University, 34450 Sar\i yer, Istanbul, T\"urkiye}
	\affiliation{T\"UBITAK Research Institute for Fundamental Sciences (TBAE), 41470 Gebze, T\"urkiye}
	
 \date{\today} 
%**************************************************************%
\begin{abstract}
Quantum thermometry has long promised sensitivities beyond classical limits, with recent paradigms highlighting quantum coherence as a primary driver of precision. While most studies focus on single-point temperature estimation, we extend this framework to the spatial domain. We consider a probe two-level system (TLS) coupled asymmetrically to an auxiliary TLS that equilibrates with a sample of profile $T(x)$. The reduced probe state carries coherence in its local basis that remains temperature-sensitive after the probe transition has frozen out, yielding a Fisher information exponentially larger at low temperature than that of a population-only readout. An exact reparametrization converts the thermometric precision into a local spatial distinguishability bound; its validity requires only a sufficiently small relative temperature uncertainty and yields an explicit measurement budget. Adding auxiliary TLSs shifts the array's optimal working temperature upward and broadens the useful sensing window, with the achievable gain ultimately limited by the total available coupling rather than by an indefinite scaling law. Finally, we propose a microwave Mach–Zehnder interferometer that maps the temperature-dependent probe coherence onto an experimentally accessible quadrature observable.
\end{abstract}
\maketitle
%**************************************************************%
\section{Introduction}

Precise control and monitoring of temperature variations at the nanoscale represent a cornerstone challenge across modern science, from probing biological processes to optimizing quantum technologies~\cite{Kucsko2013,D3NR00343D,Menges2016,Bai2016,Aspuru-Guzik2012,RevModPhys.86.153,BENENTI20171,FLEMING201138}. While bulk thermometry provides a reliable macroscopic average, it often fails to capture the underlying physics of heat transport. Spatially resolved measurements are, therefore, essential for uncovering local dissipation hotspots, heat-flow pathways, and internal gradients that remain hidden in the thermodynamic limit~\cite{Cahil2003,YANG2022107553}.

Traditional sensing modalities—including resistance thermometry~\cite{pavese2012modern}, micro-bolometers~\cite{Rieke2002}, and scanning thermal microscopy~\cite{Majumdar1999,Gomes2015}—have achieved remarkable precision. However, these classical techniques are frequently constrained by their invasive nature. Physical contact can perturb the local thermal environment, while stochastic fluctuations and the finite sampling footprint of the probe impose fundamental limits on resolution.

Quantum thermometry has emerged as a powerful paradigm to surpass these classical boundaries~\cite{Mehboudi_2019,DePasquale2016,Potts2019fundamentallimits}. By exploiting nonclassical resources such as energy-level quantization~\cite{PhysRevLett.114.220405}, quantum coherence~\cite{PhysRevResearch.5.043184, PhysRevA.110.032605, David2013, Frazao2024, Ullah_2025}, and entanglement~\cite{PhysRevA.91.012331,PhysRevResearch.2.033498,TIAN20171,PhysRevA.101.032112}, quantum sensors can maximize sensitivity as dictated by the quantum Cram\'er–Rao bound~\cite{PhysRevLett.127.190402, Mukherjee2019,DePasquale2018}. Recent progress has leveraged qubit-based probes~\cite{Bloch2012, PhysRevA.86.012125,PhysRevA.84.032105, PhysRevLett.118.130502,PhysRevA.98.042124, PhysRevA.99.062114,PhysRevResearch.2.033497, PhysRevResearch.2.033394, PhysRevA.111.062201,PhysRevLett.134.213602, ypp6-5bt9} across platforms including cold atoms~\cite{PhysRevLett.127.113602,PRXQuantum.3.040330, PhysRevA.93.043607,PhysRevLett.125.080402}, optomechanical resonators~\cite{Purdy2017}, nitrogen-vacancy centers in diamond~\cite{David2013},  and collective probes exhibiting Heisenberg scaling under idealized conditions~\cite{PhysRevA.110.062406,PhysRevApplied.16.064026,Zhang2022, Napolitano2011,Abiuso_2024}. 

Despite these advances, most current models rely on the simplifying assumption that the sample temperature $T$ is spatially uniform. This reduces the problem to a point-estimation task. In practice, however, systems such as biological tissue~\cite{KIM2015129} and superconducting circuits~\cite{Moore2014,Halbertal2016,PhysRevLett.121.157701} exhibit complex, non-uniform temperature distributions $T(x)$. Resolving such profiles presents a fundamental challenge: how can we enhance thermometric precision for spatially varying distributions? Furthermore, quantum metrology strategies can be applied to identify optimal measurements, yielding the best precision and accuracy.  

In this work, we address this challenge by investigating a probe two-level system (TLS) coupled to an auxiliary TLS, which interacts directly with a thermal bath characterized by a spatial temperature profile $T(x)$. This configuration generates steady-state coherences in the reduced state of the probe. We demonstrate that this coherence carries significant temperature dependence in the local basis, enhancing the quantum Fisher Information (QFI), surpassing what can be extracted solely from probe state populations. We validate this approach by proposing a realistic experimental readout based on microwave interferometry and homodyne detection to measure local temperature gradients. 

Our starting point is the established result that coherence generated by asymmetric probe--auxiliary coupling can substantially enhance low-temperature thermometric precision, with extensions to multiple ancillas and coherence-assisted qubit thermometry demonstrating the broader relevance of this mechanism~\cite{PhysRevResearch.5.043184,e27020204,PhysRevA.110.032605}. Here, rather than revisiting this mechanism, we focus on its implications for spatial thermometry. Specifically, we formulate thermometric precision as a spatial quantum-statistical metric with an explicit validity domain, derive a closed collective-sector description that identifies coupling-induced gap renormalization as the origin of the finite-size enhancement, and demonstrate a concrete interferometric readout scheme.

The paper is organized as follows. Section~\ref{model} introduces the probe–auxiliary model based on asymmetric coupling for coherence generation and the spatial resolution of the one-dimensional temperature profiles. We present our results in Sec.~\ref{results}, whereas in Sec.~\ref{MZI_readout}, we propose an interferometric readout based on a Mach--Zehnder interferometer. Finally, our conclusions are summarized in Sec.~\ref{conclusion}. We provide the derivations underlying the interferometric readout scheme in Appendix~\ref{Appendix_A}.

%************************************************************************%
\section{Thermometric protocol}\label{model}
\subsection{Model}

\begin{figure}[t]
	\centering
	\includegraphics[width=.9\columnwidth]{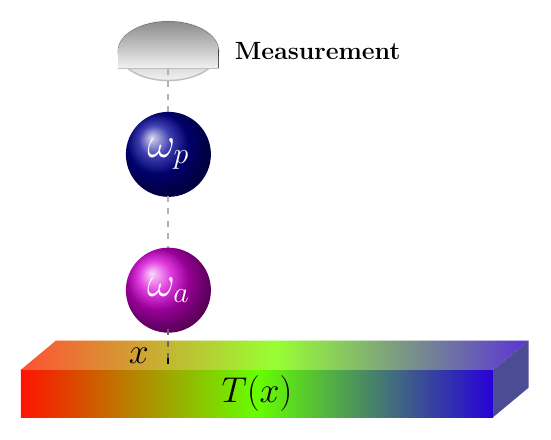}
	\caption{Scheme of the temperature sensing protocol.}
	\label{fig:scheme}
%%%%%%%%
\end{figure}

In this section, we develop a theoretical model designed to enhance the thermal sensitivity of a TLS probe through the controlled generation of coherence. As sketched in Fig.~\ref{fig:scheme}, an auxiliary TLS with frequency $\omega_{a}$ is placed in thermal contact with a sample whose temperature profile $T(x)$ varies in space. This auxiliary TLS is coupled to a primary probe TLS, characterized by transition frequency $\omega_{p}$, on which the measurement is actually performed: the sample-induced dynamics are transduced from the auxiliary to the probe through this coupling, without ever requiring the sample's full Hamiltonian to be modeled explicitly.

The interaction between the probe and the auxiliary is governed by the asymmetric Hamiltonian~\cite{PhysRevResearch.5.043184} 
\begin{equation}
	\label{model1}
	H = \tfrac{1}{2}  \hbar \omega_p  \sigma_z^{(p)} + \tfrac{1}{2} \hbar \omega_a  \sigma_z^{(a)} + \hbar g\, \sigma_z^{(a)} \sigma_x^{(p)} \, ,
\end{equation}
where $\sigma_z$ and $\sigma_x$ are the Pauli matrices of each TLS.  This Hamiltonian belongs to the same family used in earlier quantum thermal-management proposals, where two qubits coupled through an engineered bath interface were shown to yield heat-diode and heat-transistor functionality by controlling the direction and magnitude of heat flow between two reservoirs~\cite{PhysRevE.99.042121,PhysRevResearch.2.033285}. Here, we repurpose this architecture toward a different end: rather than asking how much heat flows and in which direction, we ask how much information about the bath temperature is encoded in the probe's quantum state. This shift from thermal control to quantum metrology is what motivates the asymmetric coupling: it is precisely the mechanism that seeds coherence in the probe state, and, as we show below, this coherence is what keeps the QFI from vanishing exponentially in the low-temperature regime, where single-probe thermometry limited by the probe own gap would otherwise freeze out.

%%%%%%%%

The Hamiltonian~\eqref{model1} can be diagonalized by means of a unitary transformation to the dressed basis; namely, 
\begin{equation}
	\label{H_diag}
H_{\mathrm{diag}} =  {U}  \, H  \, {U}^\dagger = \tfrac{1}{2}  \hbar \tilde{\omega}_p  \sigma_z^{(p)} + \tfrac{1}{2} \hbar \omega_a \sigma_z^{(a)} \, ,
\end{equation} 
where $\tilde{\omega}_p=\sqrt{\omega_p^2+4g^2}$ and the unitary operator $U$ is 
\begin{equation} 
	\label{uni}
{U}=\exp \left [- i \frac{\varphi}{2}  \,  \sigma^{(a)}_z\sigma_y^{(p)}\right ]=\cos \left ( \frac{\varphi}{2} \right) \openone - i \, \sigma^{(a)}_z\sigma_y^{(p)}\sin \left ( \frac{\varphi}{2} \right ) \, , 
\end{equation} 
with mixing angle $\varphi=\arctan(2g/\omega_p)$. 

We assume weak coupling of the compound probe--ancilla system to a thermal reservoir through the ancillary degrees of freedom alone, and describe its relaxation with a Markovian master equation. Because the probe--ancilla coupling $g$ is not assumed small compared to the dissipative rates, the dissipator must be built in the eigenbasis of the full interacting Hamiltonian $H_{\mathrm{diag}}$ rather than in the bare local eigenbases of probe and ancilla separately. We therefore employ a global, energy-eigenbasis master equation satisfying the Kubo--Martin--Schwinger (KMS) condition~\cite{breuer2002theory,PhysRevA.98.052123,PhysRevE.99.042121,PhysRevE.90.022102}. Under the further ergodicity assumption that no subspace remains dynamically invariant, its stationary state is the Gibbs state of the complete interacting Hamiltonian~\cite{Spohn1977},
\begin{equation}
	\label{GTS}
\tilde{\varrho}_{ap}(T)=
\frac{e^{-\beta H_{\mathrm{diag}}}}{\Tr ( e^{-\beta H_{\mathrm{diag}}}) }=\tilde{\varrho}_a(T)\otimes\tilde{\varrho}_p(T) \, ,
\end{equation}
with $\beta = 1/k_{\mathrm{B}} T$. Since $H_{\mathrm{diag}}$ is, by construction, diagonal and noninteracting in the dressed basis, its Gibbs state factorizes exactly, with no coherence or correlation whatsoever between probe and ancilla in that basis: this factorization is what lets us analyze the two subsystems independently while retaining all thermal information transmitted from the sample. Crucially, however, the dressed basis is \emph{not} the local basis in which the probe is actually prepared and read out. The bare probe operator $\sigma_z^{(p)}$ does not commute with the original Hamiltonian $H$, so the true energy eigenbasis of the compound system is the dressed basis reached by the rotation $U$ of Eq.~\eqref{uni}, not the bare local one. The probe coherence driving the enhanced sensitivity derived below is therefore not introduced by hand: it is the local-basis shadow of a state that is thermal and coherence-free in the dressed basis, becoming visible only once $\tilde{\varrho}_{ap}(T)$ is rotated back to the frame in which the probe is measured.

Rotating back to the local basis, $\varrho_{ap}(T) =   {U}^\dagger(\varphi) \, \tilde{\varrho}_{ap}(T)\,  {U}(\varphi)$, and tracing out the auxiliary TLS yields the reduced probe state:
\begin{equation}
	\label{dm1}
	\varrho_p (T) =
	\begin{pmatrix}
		p(T) & c(T) \\
		c(T) & 1 - p(T)
	\end{pmatrix} \, ,
\end{equation}
with population $p(T)$ and induced coherence $c(T)$ given by
\begin{equation}
	\begin{aligned}
	\label{pop+coh}
	p(T) & = \frac{1}{2} (1 - \tanh \vartheta_p  \cos \varphi ) \, ,  \\
	c(T) & = \frac{1}{2}  \tanh \vartheta _p \tanh \vartheta_a \sin \varphi   \, ,
	\end{aligned}
\end{equation}
where we have introduced the thermal variable $\vartheta= {\hbar \omega}/{(2k_{\mathrm{B}} T )}$ for both $\omega_{a}$ and $\tilde{\omega}_{p}$. Since the auxiliary TLS remains in a purely thermal state, we have
\begin{equation}
	\varrho_a (T) =  \begin{pmatrix}
		\tfrac{1}{2}(1-\tanh \vartheta_a)	 & 0 \\
		0  & \tfrac{1}{2}(1+ \tanh \vartheta_a )	
	\end{pmatrix} \, .
	\label{eq:rhoa}
\end{equation}

\subsection{Ultimate resolution limits}\label{UPL}

Since $p(T)$ and $c(T)$ jointly carry the temperature information, thermometry becomes a problem of quantum parameter estimation, governed by the quantum Fisher information (QFI), a mathematical measure of the sensitivity of an observable quantity to changes in its underlying parameters. It is defined as
\begin{equation}
	\mathcal{F}_{\varrho} (T)  = \Tr [ \varrho (T) \,  \mathcal{L}_{T}^{2} ] \, ,
\end{equation}
where the symmetric logarithmic derivative $\mathcal{L}_{T}$ is the self-adjoint operator satisfying the implicit equation~\cite{Helstrom1969}
\begin{equation}
	\partial_T  \varrho (T) = 	\tfrac{1}{2}  \{ \mathcal{L}_{T} , \varrho(T)  \}  \, ,
\end{equation}
where $\{\cdot, \cdot\}$ stands for the anticommutator.

For a TLS it is convenient to use the Bloch representation of the density matrix, namely
\begin{equation}
	\varrho = \tfrac{1}{2} (\openone + \mathbf{r} \cdot \bm{\sigma} ) \, ,
\end{equation}
where $\mathbf{r}$ is the Bloch vector and $\bm{\sigma}$ are the Pauli matrices. Conversely, the Bloch components are just expectation values of the Pauli matrices, $\mathbf{r}  = \Tr ( \varrho \, \bm{\sigma} )$. In this parametrization, the QFI can be computed as~\cite{PhysRevA.87.022337}
\begin{equation}
	\label{QFI}
	\mathcal{F} _{\varrho} (T) = |\partial_T \mathbf{r}|^2 + \frac{(\mathbf{r} \cdot \partial_T \mathbf{r})^2}{1 - |\mathbf{r}|^2}  \, .
\end{equation}
The first term captures sensitivity arising from changes in populations, while the second term captures contributions from coherence. A direct calculation shows that for our model
\begin{widetext}
\begin{equation}
\mathcal{F}_{\varrho_p}  (T)  = \frac{1}{T^2}\left\{
\frac{[ (C^2+S^2\tanh^2\vartheta_a )\vartheta_p\, \sech^2\vartheta_p	+ S^2\,\vartheta_a \tanh\vartheta_p \tanh\vartheta_a\,\sech^2\vartheta_a ]^2}
{ (C^2+S^2 \tanh^2\vartheta_a)  [1-\tanh^2\vartheta_p (C^2+ S^2\tanh^2\vartheta_a ) ]} +
\frac{S^2 C^2\,\vartheta_a^2\,\tanh^2\vartheta_p\,\sech^4\vartheta_a}{C^2+S^2 \, \tanh^2\vartheta_a}
\right\} \, ,
\label{QFI_zx}
\end{equation}
\end{widetext}
where $C= \cos \varphi$, $S =\sin\varphi$, and we have taken into account that  $\partial_T  \tanh \vartheta_{a} = - \vartheta_{a}/T \sech^{2} \vartheta_{a}$.  

To isolate the metrological role of coherence, we benchmark $\varrho_p$ against the state obtained by fully dephasing it in the probe $\sigma_z$ basis, so the off-diagonal coherences are removed while leaving the populations of Eq.~\eqref{pop+coh} untouched, viz
\begin{equation}
	\label{dm2}
	\bar{\varrho}_p   =
	\begin{pmatrix}
		\tfrac{1}{2}(1-\tanh \vartheta_p \cos\varphi) & 0 \\
		0 & \tfrac{1}{2}(1+\tanh \vartheta_p   \cos\varphi )
	\end{pmatrix} \, .
\end{equation}
The dephased QFI evaluates to
\begin{equation}
		\mathcal{F}_{\bar{\varrho}_p}  (T)   =  \frac{1}{4T^{4}} \frac{\tilde{\omega}_p^2 \, C^2(1- \tanh^2 \vartheta_p)^2}{1 - C^2  \tanh^2 \vartheta_p} \, .
\label{QFI_zx_d}
\end{equation} 
Obviously, the dephasing map is a completely positive trace-preserving channel that is independent of temperature, hence $ \mathcal{F}_{ \varrho_p}  (T)  \geq \mathcal{F}_{\bar{\varrho}_p}  (T)$, so that any gap between the two directly quantifies the metrological advantage supplied by coherence. 

The time-honored quantum  Cram\'er-Rao bound (QCRB)~\cite{Cramer:1946aa,Rao:1945aa} links the QFI $ \mathcal{F}_{ \varrho_p}  (T) $ with the ultimate bound achievable by the variance of any arbitrary unbiased estimator $\hat{T}$, built from $\nu$  identical and independent probes and
measurements:
\begin{equation}
	\delta \hat{T}  \ge \frac{1}{\sqrt{ \nu \; \mathcal{F}_{\varrho_p} (T) }} \, .
	\label{eq:CRB_T}
\end{equation}
The next step of the protocol connects this thermometric precision to local spatial discrimination along a calibrated temperature profile. Treating the probe as a local quantum thermometer coupled to a one-dimensional bath with profile $T=T(x)$, its sensitivity at a point $x$ follows from the fact that, under a differentiable change of variable, the QFI transforms as
\begin{equation}\label{QFI_general}
\mathcal{F}_{\varrho_p} (x) = [T^\prime(x)]^2 \, \mathcal{F}_{\varrho_p} (T) \, ,
\end{equation} 
so  the corresponding QCRB for  the spatial coordinate reads
\begin{equation}\label{delta_x}
\delta \hat{x} \ge \frac{1} {|T^\prime(x)|\sqrt{\nu\; \mathcal{F}_{\varrho_p} (T)}} = \frac{1}{\sqrt{\nu\;\mathcal{F}_{\varrho_p} (x)}} \, .
\end{equation}
Two nearby points $x$ and $x+\Delta x$ are thermometrically distinguishable only  $ | T(x+\Delta x) - T(x) |  \ge  \delta \hat{T}$, which for a smooth profile is linearized as $ | T(x+\Delta x)-T(x) | \simeq  | \partial_x T(x) |  \,  | \Delta x | $.  Thus, the local spatial distinguishability bound is
\begin{equation}
\Delta x_{\mathrm{loc}} = \frac{1} {|T'(x)|\sqrt{\nu\,\mathcal{F}_{\varrho_p} (T)}}.
\label{eq:spatial_resolution_final}
\end{equation}
The elementary linearization argument therefore reproduces Eq.~\eqref{delta_x} exactly. What Eq.~\eqref{QFI_general} adds is not a different number but a different status: the bound follows from an exact reparametrization of the quantum statistical model rather than from error propagation, and its domain of validity is correspondingly explicit.
Equivalently, the Bures line element~\cite{Bures:1969aa} can be written as
\begin{equation}
ds_B^2 = \frac{1}{4} \mathcal{F}_{\varrho_p} (T) \,dT^2 = \frac{1}{4}\mathcal{F}_{\varrho_p} (x) \,dx^2,
\end{equation}
showing that the temperature profile pulls the thermometric quantum-statistical metric back onto the spatial coordinate. The local bound $\Delta x_{\mathrm{loc}}(x)$ shows that spatial discrimination depends jointly on two factors: the local $\mathcal{F}_{\varrho_p} (T)$, which quantifies intrinsic temperature sensitivity, and the temperature gradient $|T'(x)|$. Enhancements in the QFI --- for instance, through coherence, entanglement, or optimized measurement strategies --- therefore translate directly into improved local spatial discrimination.

Importantly, this conclusion does not rely on any particular form of the temperature profile; it applies generally to any sufficiently smooth function~$T(x)$. Regions where the temperature slowly varies  ($|\partial_x T(x)| \ll 1$) impose strict limitations on spatial discrimination, while steep gradients amplify the resolving power of the probe. Equation~\eqref{eq:spatial_resolution_final} can be interpreted as a \textit{local spatial distinguishability bound}, relating temperature sensitivity to spatial resolution for a known profile $T(x)$. Its validity relies on standard assumptions: local thermal equilibrium, a probe footprint much smaller than the temperature-variation length scale $\ell_{\rm probe}\ll\ell_T\equiv T/|T'(x)|$, negligible thermometric back action, a locally monotonic (invertible) temperature profile, and the self-consistency condition $\Delta x_{\mathrm{loc}}\ll\ell_T$. Under these conditions, the probe acts as an effective local sensor, and the spatial resolution is directly set by the thermometric precision encoded in $ \mathcal{F}_{\varrho_p} (T)$. 

%**************************************************%
\section{Results}
\label{results}
%**************************************************%

\subsection{Quantum Fisher information analysis}

We now examine the precision of temperature estimation and the resulting spatial resolution using the QFI derived above. Throughout, we fix $\omega_p = 1.0$ and $\omega_a = 0.04$, so all quantities are expressed in units of $\omega_p$ and reported without explicit units from here on.

\begin{figure}[t]
	\centering
	\includegraphics[width=.95\linewidth]{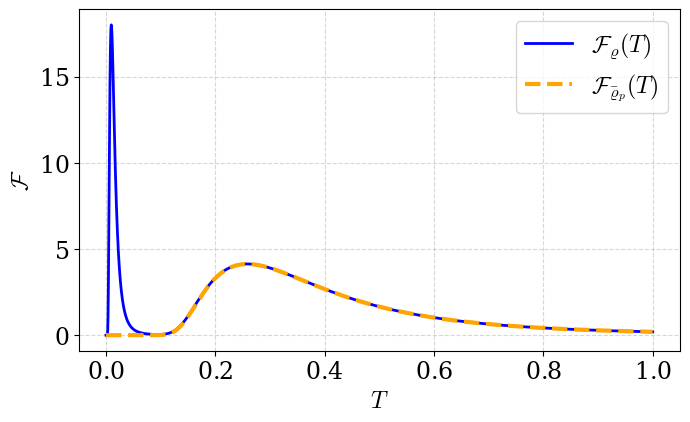}
	\caption{QFI for full state $\varrho_p (T)$ (dot-dashed blue) and dephased state $\bar{\varrho}_{p}(T))$ (dashed red) on the points $x$ of the sample where the temperature is $T=T(x)$. The rest of the parameters are set to $g=0.04$, $\omega_p=1$, $\omega_a=0.04$, and $\nu=1$.}
	\label{fig:comp}
\end{figure}

Figure~\ref{fig:comp} plots the analytical results of Eqs.~\eqref{QFI_zx} and \eqref{QFI_zx_d}: the dot-dashed blue curve shows the QFI of the full state $\varrho_p$, which sets the ultimate quantum limit on estimation precision, while the dashed red curve shows the dephased counterpart $\bar{\varrho}_p$. Besides the expected high-temperature peak, $\mathcal{F}_{\varrho_p} (T) $ develops a second, larger peak at low temperature—one that grows further with increasing coupling strength $g$ and is absent from the dephased curve.

This feature reflects a competition between two distinct energy scales: the dressed-probe gap $\hbar\tilde\omega_p$ and the auxiliary gap $\hbar\omega_a$, with $\omega_a \ll \tilde\omega_p$. Near the low-temperature peak, a direct  analytical approximation gives
\begin{equation}
	\label{approx_QFI_full}
	\mathcal{F}_{\varrho_p} (T)  \approx \frac{1}{T^2}  S^2 \, \vartheta_a^2 \,\sech^2\! \vartheta_a \, ,   
\end{equation}
showing explicitly that this peak is controlled by the smaller energy scale $\hbar\omega_a$; i.e., it is associated with the auxiliary TLS rather than the probe transition itself.

%----------------------------------------------------------------%
%-----------------------------------------------------------------%

The mechanism becomes transparent from the interaction Hamiltonian $H_{\mathrm{int}} = \hbar g\,\sigma_z^{(a)}\sigma_x^{(p)}$: the auxiliary state fixes the direction of an effective transverse field on the probe,
\begin{equation}
	H_p^{(\pm)}=\tfrac{1}{2} \hbar\omega_p \, \sigma_z^{(p)} \pm \hbar g\,\sigma_x^{(p)} \, ,
\end{equation}
depending on whether the auxiliary sits in the $\sigma_z^{(a)}=\pm1$ eigenstate. The resulting field lies in the $xz$ plane with magnitude $\hbar\tilde\omega_p$ and tilt angle $\varphi$ ($\sin\varphi = 2g/\tilde\omega_p$, $\cos\varphi=\omega_p/\tilde\omega_p$), so $\hbar g$ sets how efficiently the auxiliary state imprints coherence onto the probe.

Because the auxiliary is directly coupled to the bath, its population imbalance $\vartheta_a$ tracks the bath temperature. In the regime $k_BT \ll \hbar\tilde{\omega}_p$, where the probe is already frozen ($\tanh \vartheta_p \simeq1$), the induced probe coherence reduces to $r_x \simeq \sin\varphi \,\tanh \vartheta_a$: the probe coherence becomes a faithful proxy for the auxiliary thermal polarization. Meanwhile, the probe population is essentially temperature-independent here, since $\partial_T \tanh \vartheta_p  \sim e^{- \vartheta_p}$ is exponentially small—which is precisely why the dephased state shows no low-temperature peak.  This defines a transduction chain \mbox{$\delta T \mapsto \delta \vartheta_a \mapsto \delta r_x$}, whereby a temperature change alters the auxiliary's Boltzmann-weighted population: $\delta \tanh \vartheta_a \simeq \partial_T \tanh \vartheta_a  \,\delta T$, which is converted, through the interaction, into a shift of probe coherence $\delta r_x \simeq \sin\varphi \, \partial_T \tanh \vartheta_a  \,\delta T $. The sensitivity of this chain is thus governed by $\partial_T \tanh \vartheta_a = - \vartheta_{a}/T \sech^{2} \vartheta_{a} $: this contains a  prefactor $\vartheta_{a}/T$ that sets the overall response scale while $\sech^2(\hbar\omega_a/2k_BT)$ is the auxiliary's thermal susceptibility. At $k_BT \ll \hbar\omega_a$, the auxiliary is frozen, and the susceptibility vanishes; at $k_BT \gg \hbar\omega_a$ the susceptibility saturates, and the imbalance $\tanh \vartheta_a$ becomes small. In the regime $k_{B}T \ll \hbar \tilde{\omega}_{p}$ the probe is frozen, $\tanh \vartheta_{p} \to 1$ and $\partial_{T} \tanh \vartheta_{p} \to 0$, so only the coherence-assisted contribution survives and $\mathcal{F}_{\varrho_{p}}$ reduces to Eq.~\eqref{approx_QFI_full}. Since $\omega_{a} \ll \tilde{\omega}_{p}$, this regime still contains the window $k_{B}T \sim \hbar\omega_{a}$ in which the auxiliary remains thermally active. The  maximum of this function is determined by the condition $\vartheta_a \tanh \vartheta_{a} =2$, which gives $k_\mathrm{B} T_\mathrm{max} \approx 0.24208\hbar\omega_a$. Notice that the condition $\vartheta_a \tanh \vartheta_a = 2$ is the metrological counterpart to the classical Schottky anomaly in two-level heat capacity, where the peak occurs at $\vartheta \tanh \vartheta = 1$ ($k_B T \approx 0.417 \hbar\omega_a$)~\cite{Schlogl1985,Janyszek1986,Correa2015,DePasquale2016}. Because thermal parameter estimation links the quantum Fisher information directly to energy fluctuations via $\mathcal{F}_T = C_V / (k_B T^2)$, the additional $1/T^2$ error-propagation weighting penalizes higher temperatures, shifting the optimal sensitivity point downward to $k_B T_{\max} \approx 0.242 \hbar\omega_a$.

Finally, the peak amplitude is set by $S^2 = 4 g^2/(\omega_p^2+4g^2)$: it vanishes for $g=0$, when no auxiliary-to-probe coherence transfer occurs, and grows with increasing coupling. In short, the low-temperature peak arises because the auxiliary's smaller energy scale $\hbar\omega_a$ remains thermally active even after the probe itself has frozen out, and the interaction converts this residual sensitivity into probe coherence — extending the thermometer reach to temperatures well below the probe gap.

\subsection{Spatial resolution}
%------------------------------------------------------%
\begin{figure}[t!]
    \centering
   \includegraphics[width =\linewidth]{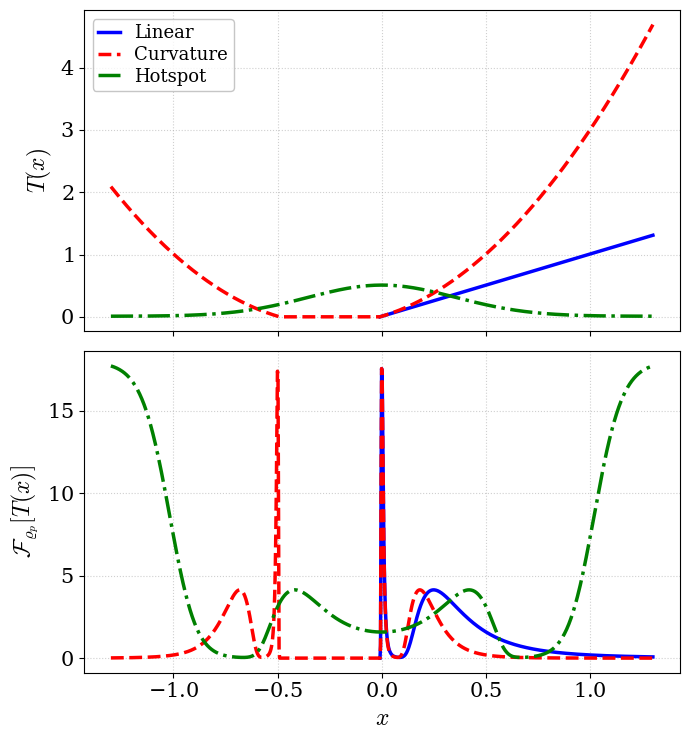}
    \caption{Top panel shows the three different profiles of $T(x)$. The bottom panel shows the corresponding QFI for the full state $\varrho_{p}$ as a function of $x$ for three different temperature profiles $T(x)$. For the linear profile, the spatial domain is restricted to $x \gtrsim -0.01$ ($T_0 = 0.01$, $\alpha = 1$) to enforce $T(x) > 0$ across the plotted range. The parameters for these profiles are set to $\alpha = 1$, $\beta = 2$, $\Delta T = 0.5$, $x_0 = 0$, $w = 0.5$, and $T_0=0.01$. The remaining system parameters are set to $\omega_p = 1.0$, $\omega_a = 0.04$, and $g = 0.04$.}
    \label{fig:QFI_vs_x}
\end{figure}
%------------------------------------------------------%
To illustrate the spatial resolution of the protocol in concrete settings, we first consider three representative temperature profiles,
\begin{flalign}
T(x) & =  T_0 + \alpha x, &   &  \text{linear gradient}, \label{eq:Tp1} \\
T(x) & = T_0 + \alpha x + \beta x^2  , &  & \text{curvature estimation}, \\
T(x)  & = T_0 + \Delta T \, e^{-(x-x_0)^2/w^2}  , &  & \text{hotspot detection} , \label{eq:Tp3}
\end{flalign}
spanning uniform gradients, curvature, and localized structure. 
The corresponding QFI $\mathcal{F}_{\varrho_p}  (x)$ are shown in Fig.~\ref{fig:QFI_vs_x}. Since $\mathcal{F}_{\varrho_p}  (T)$ is a nonlinear function of temperature, the spatial profile of $\mathcal{F}_{\varrho_p}  (x)$ inherits pronounced peaks wherever $T(x)$ crosses the probe's optimal working point $T_{\max}$, and is suppressed elsewhere. A linear gradient crosses an optimum once, while the quadratic profile can cross it twice owing to its curvature. Crucially, because $\mathcal{F}_{\varrho_p}(T)$ possesses both a low-temperature coherence-assisted maximum ($k_B T \simeq 0.242 \hbar\omega_a$) and a high-temperature resonant maximum ($k_B T \sim \hbar\tilde{\omega}_p$), a hotspot profile whose peak temperature exceeds both thresholds produces a four-peak structure in $\mathcal{F}_{\varrho_p}(x)$, corresponding to spatial crossings of both thermometric working points as the temperature rises and falls around $x_0$.

\begin{figure}[t!]
		\centering
		\includegraphics[width=\linewidth]{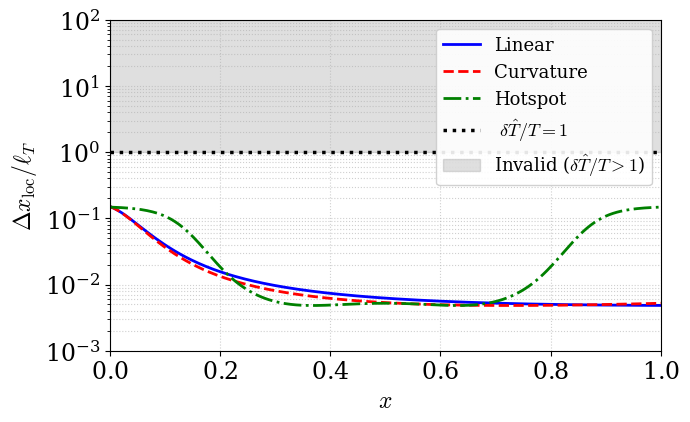}
		\caption{Local spatial distinguishability $\Delta x_\text{loc}(x)$ as a function of $x$ for three distinct thermal landscapes, such as linear gradient (solid blue), curvature estimation (dashed red), and hotspot detection (dot-dashed green). Here, $\Delta x_{\rm loc}$ denotes the $1\sigma$ spatial uncertainty (standard deviation) of the estimator; a corresponding two-sided 95\% confidence interval has a width of approximately \(4\Delta x_{\rm loc}\) for Gaussian errors. The parameters for these profiles are set to $\alpha = 0.3$, $\beta = 0.2$, $\Delta T = 0.5$, $x_0 = 0.5$, $w = 0.2$, and $T_0=0.1$. The rest of the parameters are set to $g=0.04$, $\omega_p=1$, $\omega_a=0.04$, $\nu=10^5$, and $\alpha=0.3$.}
		\label{fig:Delta_x_vs_x}
	\end{figure}

The QFI $\mathcal{F}_{\varrho_p}  (T)$ in Fig.~\ref{fig:comp} reaches its maximum value at an optimal sensing temperature $T_{\max}$. For the three temperature profiles considered, the corresponding spatial peak locations can be found as
\begin{align}
	x^*_{\text{lin}} &= \frac{T_{\max} - T_0}{\alpha},\\
	x^*_{\text{curv}} &= \frac{-\alpha \pm \sqrt{\alpha^2 - 4\beta(T_0 - T_{\max})}}{2\beta},\\
	x^*_{\text{hot}} &= x_0 \pm w \sqrt{\ln\left( \frac{\Delta T}{T_{\max} - T_0} \right)}.\label{eq:xstart}
\end{align}
Substituting Eqs.~\eqref{eq:Tp1}--\eqref{eq:Tp3} into Eq.~\eqref{eq:spatial_resolution_final} yields the local distinguishable distance $\Delta x_\text{loc}$. It is important to note that the local continuum validity condition $\Delta x_{\mathrm{loc}} \ll \ell_T(x)$ is purely thermometric. Using spatial uncertainty $\Delta x_{\mathrm{loc}}(x)$ of Eq.~\eqref{eq:spatial_resolution_final}
and the local thermal gradient length scale $\ell_T(x) = T(x) / |T'(x)|$, the spatial gradient cancels identically:
\begin{equation}
    \frac{\Delta x_{\mathrm{loc}}(x)}{\ell_T(x)} = \frac{1}{|T'(x)| \sqrt{\nu \mathcal{F}_T(T)}} \cdot \frac{|T'(x)|}{T(x)} = \frac{1}{T(x) \sqrt{\nu \mathcal{F}_T(T)}} = \frac{\delta \hat{T}}{T}.
    \label{eq:thermometric_condition}
\end{equation}
Consequently, self-consistency ($\Delta x_{\mathrm{loc}} \ll \ell_T$) simplifies to the requirement that the relative local temperature uncertainty is small ($\delta \hat{T}/T \ll 1$). This is a property of the thermometer alone and carries no profile dependence. For a target precision budget $\varepsilon \equiv \delta \hat{T} / T$, Eq.~\eqref{eq:thermometric_condition} establishes the analytical design rule for the minimum measurement budget:
\begin{equation}
    \nu_{\mathrm{min}}(x) = \left[ \varepsilon^2 T(x)^2 \mathcal{F}_T(T(x)) \right]^{-1}.
    \label{eq:nu_design_rule}
\end{equation}
At the coherence-assisted optimal working point ($T_{\mathrm{max}}$), this condition becomes parameter-free. Substituting $\mathcal{F}_{\mathrm{max}} = 4.654 S^2 / \omega_a^2$ and $k_B T_{\mathrm{max}} = 0.24208 \hbar \omega_a$, the product evaluates to $T^2 \mathcal{F}_T = 0.2727 S^2$ exactly. The relative temperature uncertainty at peak sensitivity is therefore
\begin{equation}
    \left. \frac{\delta \hat{T}}{T} \right|_{T_{\mathrm{max}}} = \frac{1.915 S}{\sqrt{\nu}} \simeq \frac{0.96 \omega_p}{g \sqrt{\nu}} \quad (g \ll \omega_p).
    \label{eq:relative_T_uncert}
\end{equation}
For parameters $g = \omega_a = 0.04$ and $\omega_p = 1$, a sample budget of $\nu = 10^4$ gives $\delta \hat{T}/T = 0.24$ at the low-temperature peak—which is below unity but not strictly $\ll 1$. To achieve a target relative error of $\varepsilon = 0.1$, Eq.~\eqref{eq:nu_design_rule} dictates $\nu \approx 5.8 \times 10^4$. Hence, we fix $\nu = 10^5$ across our spatial resolution plots, ensuring strict validity without empirical fitting. Figure~\ref{fig:Delta_x_vs_x} shows that $\Delta x_\text{loc}$ for the linear profile (solid blue) sets a baseline governed by the local slope; the quadratic profile (dashed red) follows a similar trend, modulated by curvature. The hotspot profile (dash-dotted green) behaves qualitatively differently, with degraded resolution far from the hotspot and sharp enhancement near it — reflecting how strongly the local effective gradient, and hence the resolving power, depends on probe position there.

For the linear case, Eq.~\eqref{eq:spatial_resolution_final} reduces to the transparent benchmark
\begin{equation}
	\Delta x_{\mathrm{loc}} = \frac{1}{ | \alpha |  \sqrt{\nu \; \mathcal{F}^{\mathrm{lin}}_{\varrho_p}  (x)}} \, ,
		\label{eq:dx_linear_case}
	\end{equation}
making explicit that steeper gradients and larger QFI both sharpen spatial resolution. For richer profiles, resolution is no longer set by the local gradient alone but by how precisely parameters such as curvature $\beta$ or hotspot position $x_0$ and width $w$ can be estimated~\cite{deshler2025}—turning thermal imaging into a multiparameter estimation problem governed by the QFI matrix~\cite{Szczykulska03072016}. In practice, the probe also samples a finite spatial region, which further limits resolution and smooths fine profile features. 

%----------------------------------------------------------%
\begin{figure}[t!]
    \centering
    \includegraphics[width=\linewidth]{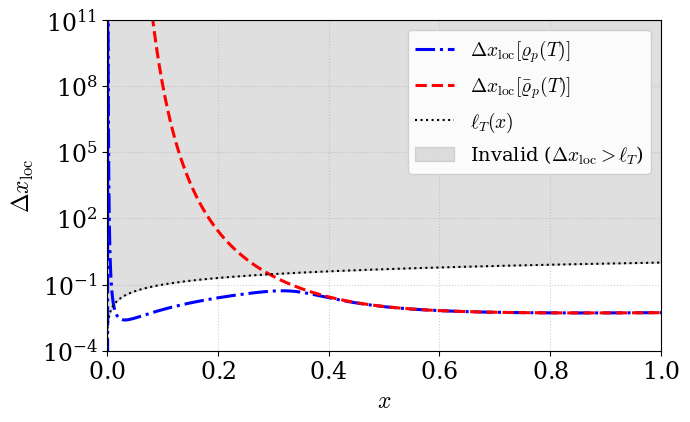}
    \caption{Local spatial distinguishability $\Delta x_\text{loc}$ of Eq.~\eqref{eq:dx_linear_case} at each point of sample profile $x$ where the temperature is $T(x)=T_0+\alpha x$ is linear gradient. The parameters are set to $\omega_p=1$, $\omega_a/\omega_p=0.04$, $g=0.04$, $\alpha=0.3$, $T_0=0$, and $\nu=10^5$. As $x\to0$, also $T\to0$ and consequently $\mathcal{F}_{\bar{\varrho}_p}(T)\to 0$; therefore, the probe has completely lost local resolving power there at any unit scale $[L]$, driving the relative temperature uncertainty $\Delta x_\text{loc}$ to diverge.}
    \label{fig:Delta_x_full_reduced}
\end{figure}

%----------------------------------------------------------%
Figure~\ref{fig:Delta_x_full_reduced} compares $\Delta x_{\mathrm{loc}}$ for the full and dephased probe states. Coherence keeps $\Delta x_\text{loc}$ small throughout, whereas its removal degrades resolution substantially; increasing the coupling strength $g$ further sharpens the quantum result. For an ensemble budget of $\nu = 10^5$, the coherent probe maintains operational validity ($\Delta x_{\mathrm{loc}} \le \ell_T$) down to $x \approx 0.01$, whereas the dephased probe breaks down much earlier at $x \approx 0.35$.  Quantum coherence thus expands the usable spatial operating range of the sensor by a factor of roughly $35$. 
Taking the low-temperature limit in Eqs.~\eqref{QFI_zx_d} and \eqref{approx_QFI_full}, the QFI expressions for the respective sectors reduce to
\begin{equation}
    \mathcal{F}_{\bar{\varrho}_p} \simeq \frac{4 C^2 \tilde{\omega}_p^2}{S^2 T^4} e^{-2\tilde{\omega}_p / T}, \qquad 
    \mathcal{F}_{\varrho_p} \simeq \frac{S^2 \omega_a^2}{T^4} e^{-\omega_a / T},
\end{equation}
yielding the ratio of sensitivities
\begin{equation}
    \frac{\mathcal{F}_{\varrho_p}}{\mathcal{F}_{\bar{\varrho}_p}} \simeq \frac{S^4}{4 C^2} \frac{\omega_a^2}{\tilde{\omega}_p^2} \exp \left[ \frac{\hbar (2\tilde{\omega}_p - \omega_a)}{k_B T} \right].
\end{equation}
At $\omega_p = 1$, $\omega_a = g = 0.04$, and $T = 0.03$, this evaluates to $\Delta x^{\text{deph}} / \Delta x^{\text{coh}} = 2.4 \times 10^{10}$, in excellent agreement with the numerical value of roughly $10^{10}$ observed at $x \simeq 0.1$ in Fig.~\ref{fig:Delta_x_full_reduced}. Figures~\ref{fig:Delta_x_vs_x} and~\ref{fig:Delta_x_full_reduced} present the spatial resolution profiles evaluated directly at a measurement budget of $\nu = 10^5$, ensuring the local continuum condition $\Delta x_{\mathrm{loc}}(\nu) = \Delta x_{\mathrm{loc}}/\sqrt{\nu} \ll \ell_T(x)$ is satisfied throughout the valid domain. Physically, the rate-limiting step for this budget is the auxiliary system's re-equilibration with the thermal sample.  For representative circuit-QED parameters ($\omega_p/2\pi = 5~\text{GHz}$ and coupling $g = 10^{-2}\omega_p$), each equilibration cycle takes $\sim 3~\text{ns}$, allowing the full $\nu = 10^5$ budget to accumulate in just $\sim 0.3~\text{ms}$—well within standard experimental integration times.

%----------------------------------------------------------%
%----------------------------------------------------------%

To connect this intrinsic sensitivity to the effect of dephasing, we define the \emph{dephasing susceptibility of the thermometric information} as the fractional total QFI
lost upon dephasing
\begin{equation}
	\eta_{\mathrm{coh}}(x) 	=1- \frac{\mathcal{F}_{\bar{\varrho}_p}(T)}	{\mathcal{F}_{\varrho_p}(T)} 	\in[0,1] \, .
\end{equation}
This quantity measures the fraction of thermometric information lost when the probe coherence is removed; it is not a measurement efficiency, since no measurement process enters its definition. In the low-temperature limit, the population-only QFI of the dephased probe vanishes exponentially as $T\to0$, while the undephased probe can retain substantial temperature sensitivity. Consequently, $\eta_{\mathrm{coh}}(x)\to1$ as $T\to0$, indicating that an increasingly large fraction of the available thermometric information is coherence-enabled.

\begin{figure}[t]
    \centering
    \includegraphics[width=\linewidth]{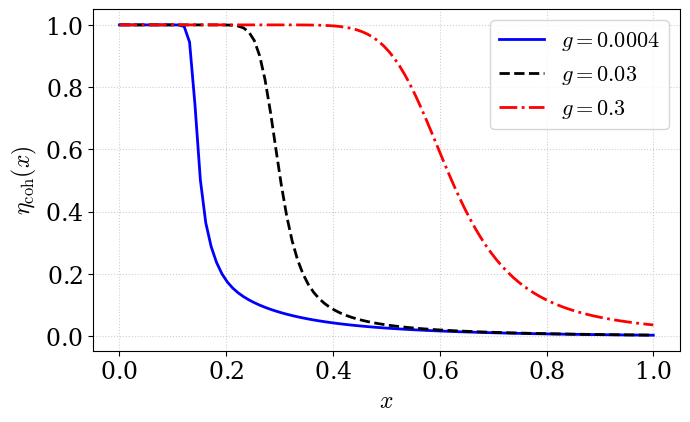}
    \caption{Dephasing susceptibility of the thermometric information $\eta_\text{coh}(x)$ as a function of $x$. Here, the solid blue, dashed black, and dot-dashed red curves correspond to $g=4\times10^{-4}$, $g=0.03$, and $g=0.3$, respectively. The remaining parameters are set to $\omega_p=1$, $\omega_a=0.04$, $\nu=1$, $T_0=0.01$, and $\alpha=0.3$.}
    \label{fig:R}
\end{figure}

Figure~\ref{fig:R} shows the fraction $\eta_{\mathrm{coh}}(x)$ as a function of $x$ across a linear temperature gradient for varying coupling strengths $g$. As $x \to 0$ ($T \to T_0$), $\eta_{\mathrm{coh}}(x)$ approaches unity across all coupling strengths, demonstrating that classical population sensitivity freezes out at low temperatures, leaving quantum coherence as the sole carrier of temperature information. Increasing the coupling $g$ drastically extends the coherence-dominated regime ($\eta_{\mathrm{coh}} \approx 1$) toward higher spatial positions $x$, reinforcing that strong probe-auxiliary interaction is critical for preserving quantum-enhanced thermometric precision away from absolute zero.

%------------------------------------------------------------------------%
%--------------------------------------------------------------------%
\subsection{Scaling analysis for multiple auxiliary TLSs}
In this section, we investigate how the local spatial distinguishability scales with the number of auxiliary TLSs. This situation is described by the Hamiltonian
\begin{equation}
H= \tfrac{1}{2} \hbar \omega_p \sigma_z^{(p)} + \sum_{n=1}^{N}\tfrac{1}{2} \hbar \omega_n \sigma_z^{(n)} + \hbar g\, \sum_{n=1}^{N}\sigma_z^{(n)} \sigma_x^{(p)},
\end{equation}
with all auxiliary frequencies set equal, $\omega_n=\omega$. Assuming weak coupling to a common bath, the full system, probe included, equilibrates to~\cite{Spohn1977}
\begin{equation}
    \varrho_T = \frac{1}{\mathcal{Z}} e^{-\beta H},
\end{equation}
where $\mathcal{Z}=\Tr (e^{-\beta H})$ is the partition function.   We introduce $ m = \sum_{n=1}^{N} s_n$, with  $ m = -N, -N+2, \dots, N$,
and  degeneracy $\binom{N}{(N+m)/2}$. For fixed $m$, we have
\begin{equation}
H_p^{(m)} = \tfrac{1}{2} \hbar \omega_p \sigma_z +  \hbar g_N m \sigma_x \, ,
\end{equation}
where $\Omega_m = \sqrt{\omega_p^2 + 4g_N^2 m^2}$ with $g_{N} \in \{g, g/N, gN\}$ denotes the resource-scaling law, so the three curves of Fig.~7 correspond to fixed per-ancilla coupling, fixed total coupling, and growing total coupling, respectively. The partition function is
\begin{equation}
\mathcal{Z}_N = 2 \sum_{m} d_m e^{- m \vartheta_a} \cosh  ( \vartheta_m) \, 
\end{equation}
and $\vartheta_m$ is defined analogously as $\vartheta_a$, but in terms of the frequency $\Omega_m$.  The reduced probe Bloch components can be evaluated exactly as
%--------------------------------------------------------------%
\begin{figure*}[t!]
    \centering
    \includegraphics[width=0.32\linewidth]{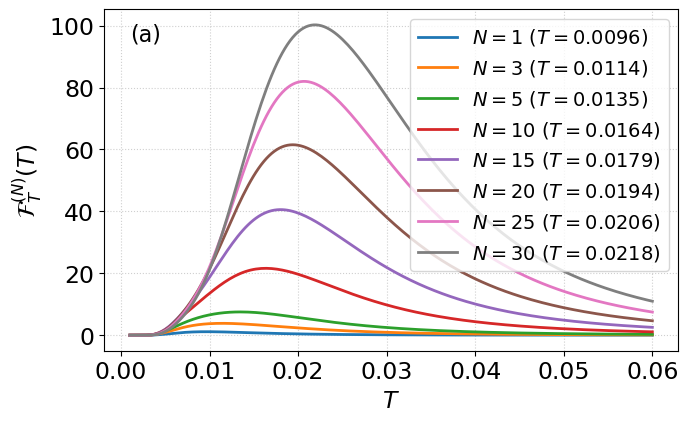}
        \includegraphics[width=0.32\linewidth]{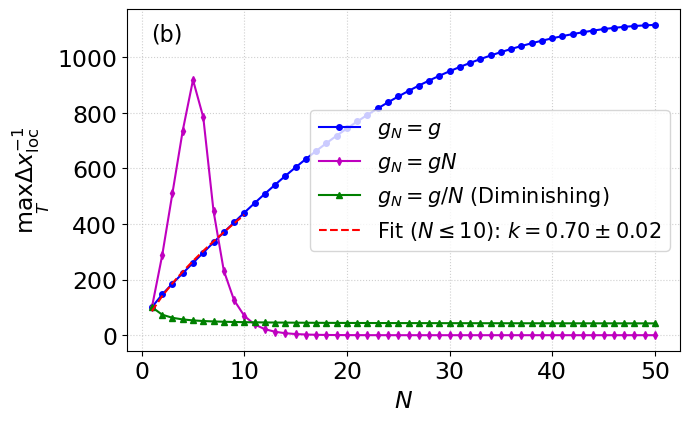}
                \includegraphics[width=0.32\linewidth]{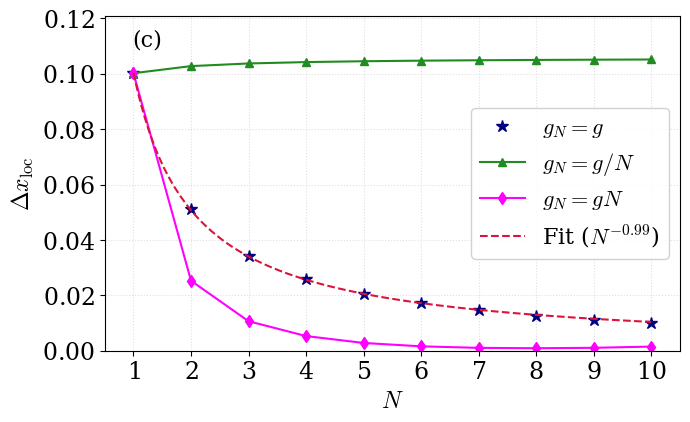}
    \caption{ Performance and scaling analysis of thermal sensing using auxiliary TLSs. (a) QFI $\mathcal{F}_T^{(N)}$ as a function of temperature $T$ for various system sizes $N$, showing the peak temperature $T_{\mathrm{max}}(N)$ shifting higher with increasing $N$. (b) Optimized spatial resolution metric $\max_T \Delta x_{\mathrm{loc}}^{-1}$ versus $N$ up to $N = 50$, demonstrating the crossover between constant ($g_N = g$), collective ($g_N = gN$), and diminishing ($g_N = g/N$) coupling regimes, alongside a power-law fit ($N \le 10$). (c) Local spatial distinguishability $\Delta x_{\mathrm{loc}}$ of a probe as a function of $N$ at a fixed local temperature $T = 0.06$. Blue star markers denote constant coupling $g_N = g$ with a power-law fit (red dashed line), green triangles represent diminishing coupling $g_N = g/N$, and magenta diamonds correspond to collective coupling $g_N = gN$. Fixed parameters: $\omega_p = 1$, $\omega_a = 0.04$, $g = 0.01$, $\alpha = 0.3$, and $\nu = 10^5$.}
    \label{fig:scaling_analysis}
\end{figure*}

\begin{equation}
	\begin{aligned}
	r_x^{(N)} & = - \frac{2 \sum_{m} \frac{2 m d_m g_N} {\Omega_m} e^{- m \vartheta_a}   \sinh \vartheta_m}{\mathcal{Z}_N}, \\
    r_z^{(N)} & = -\frac{2 \sum_{m} \frac{d_m\omega_p}{\Omega_m} e^{-m  \vartheta_a}    \sinh \vartheta_m}{\mathcal{Z}_N} .
    \end{aligned}
\end{equation}

The corresponding QFI $\mathcal{F}_T^{(N)}$ for an array of $N$ auxiliary qubits is computed using Eq.~\eqref{QFI}. To evaluate how array scaling enhances spatial resolution while remaining grounded in resource limits, we analyze the thermal spectrum dynamics before examining optimized precision metrics and fixed-temperature comparisons. First, as shown in Fig.~\ref{fig:scaling_analysis}(a), the QFI $\mathcal{F}_T^{(N)}(T)$ reveals that the auxiliary array acts as a self-tuning thermometer. In the low-temperature, two-sector regime, the thermally relevant ground sector is $m=-N$, while the first excited sector is $m=-N+2$, with degeneracy $N$. The corresponding sector splitting is
\begin{equation}
\delta_N=
\hbar\omega_a
+\frac{\hbar}{2}\left(\Omega_N-\Omega_{N-2}\right),
\label{eq:delta_N}
\end{equation}
where $\Omega_m=\sqrt{\omega_p^2+4g_N^2m^2}$. Thus, the coupling-induced renormalization of the sector gap makes $\delta_N$ explicitly dependent on the array size $N$. In the weak-coupling limit, this gives
\begin{equation}
\delta_N
\simeq
\hbar\omega_a+
\frac{4\hbar g_N^2}{\omega_p}(N-1),
\end{equation}
for fixed $g_N=g$. Since the QFI maximum in the two-sector regime occurs at a temperature set by this effective splitting, $T_{\mathrm{max}}(N)$ correspondingly shifts with $N$. As seen in Fig.~\ref{fig:scaling_analysis}(a), the optimal working temperature increases from $0.0096$ to $0.0218$ as $N$ grows from $1$ to $30$. The array therefore provides a self-tuning thermometric platform whose high-sensitivity operating point is progressively shifted to higher temperatures as the array size increases.

To account for this temperature peak tracking, Fig.~\ref{fig:scaling_analysis}(b) evaluates the maximized inverse spatial resolution metric $\max_T \Delta x_{\mathrm{loc}}^{-1}$ up to $N = 50$. The scaling behavior can be understood analytically in terms of the effective collective mixing parameter. In the weak-mixing regime, $2g_NN\ll\omega_p$, the auxiliary qubits are approximately independent, with
\begin{equation}
\langle m\rangle\simeq -N\tanh\vartheta_a,
\qquad
S_N=\frac{2g_NN}{\omega_p},
\label{eq:SN}
\end{equation}
and the probe coherence correspondingly scales as
\begin{equation}
r_x^{(N)}\simeq S_N\tanh\vartheta_a.
\end{equation}
Thus, in this regime, the $N$-auxiliary problem maps onto the corresponding single-auxiliary result through the replacement $S\rightarrow S_N$. As $N$ increases, however, this weak-mixing description breaks down when $2g_NN\sim\omega_p$, marking the onset of saturation. The associated crossover scale is therefore
\begin{equation}
N_{\mathrm{sat}}\sim\frac{\omega_p}{2g_N}.
\label{eq:Nsat}
\end{equation}
For fixed coupling, $g_N=g=0.01$ with $\omega_p=1$, this gives $N_{\mathrm{sat}}\simeq50$, while for the collective scaling $g_N=gN$ it gives $N_{\mathrm{sat}}\simeq\sqrt{\omega_p/(2g)}\simeq7$. These estimates account for the saturation of the constant-coupling enhancement near $N\sim50$ and the resonance-like maximum of the $g_N=gN$ curve near $N\sim7$ observed in Fig.~\ref{fig:scaling_analysis}(b). Over the initial fit window $N \in [1, 10]$, the optimized constant-coupling curve ($g_N = g$) obeys a power-law scaling 
\begin{equation}
    \max_T \Delta x_{\mathrm{loc}}^{-1} = A N^k,
\end{equation}
with exponent $k = 0.70 \pm 0.02$, amplitude $A = 0.3684 \pm 0.0136$, and coefficient of determination $R^2 = 0.9963$. Extending it to larger array sizes reveals that this enhancement steadily grows before beginning to saturate around $N \sim 50$ (reaching $\max_T \Delta x_{\mathrm{loc}}^{-1} \approx 3.5$), confirming that the gain at fixed $g$ is a resource-dependent, finite-size effect. Meanwhile, the collective regime ($g_N = gN$) exhibits a sharp resonance-like peak at $N = 5$ ($\max_T \Delta x_{\mathrm{loc}}^{-1} \approx 3.0$) before rapidly falling off and dropping below the constant-coupling regime near $N \approx 8$. Beyond $N \approx 13$, the collective precision vanishes as the system transitions out of the optimal parameter window, leaving the constant-coupling regime far superior for larger array sizes. Finally, the precision metric for diminishing coupling ($g_N = g/N$) decays monotonically toward a low floor near $0.18$, confirming that diluting a fixed coupling resource cannot sustain spatial precision.

Finally, Fig.~\ref{fig:scaling_analysis}(c) examines a fixed low-temperature slice at $T = 0.06$ across the different coupling prescriptions $g_N$, serving to illustrate why evaluating precision at a single operating point overstates the true scalable gain compared to the optimized metric in panel (b). For constant coupling ($g_N = g$), spatial resolution sharpens from $\Delta x_{\mathrm{loc}} \approx 0.10$ at $N=1$ down to $\approx 0.01$ at $N=10$, following a power-law fit over $N \in [1, 10]$ given by
\begin{equation}
\Delta x_{\mathrm{loc}} = \frac{C}{N^p} , \qquad C = 0.10, \quad p = 0.99,
\end{equation}
where $R^2 = 0.9999$. This fixed-$T$ improvement is ultimately limited by the saturation of the effective mixing parameter $S_N=2g_NN/\omega_p$. In particular, for constant coupling $g_N=g$, the weak-mixing scaling ceases to apply when $2g_NN\gtrsim\omega_p$, beyond which the susceptibility no longer increases according to the independent-auxiliary scaling. This saturation limits the fixed-temperature gain and further illustrates why the temperature-optimized metric $\max_T\Delta x_{\mathrm{loc}}^{-1}$ in Fig.~\ref{fig:scaling_analysis}(b) provides the more appropriate benchmark for scalable performance. For enhanced collective coupling ($g_N = gN$), spatial uncertainty drops even more rapidly, reaching $\Delta x_{\mathrm{loc}} \approx 0.0017$ by $N = 10$. Conversely, under diminishing coupling ($g_N=g/N$), distributing a fixed coupling resource weakens the interaction per TLS, causing $\Delta x_{\mathrm{loc}}$ to increase only slightly from $0.10$ to $0.105$ over $N\in[1,10]$. This near-flat behavior is consistent with Eq.~\eqref{eq:SN}, for which $S_N=2g/\omega_p$ is independent of $N$ when $g_N=g/N$; hence, the two-sector scaling predicts approximately $N$-independent spatial resolution rather than $\Delta x_{\mathrm{loc}}\propto\sqrt{N}$. The observed $\sim5\%$ variation, therefore, confirms the expected flat scaling.

%------------------------------------------------------------------%
\begin{figure}[t]
	\centering
	\includegraphics[scale=0.40]{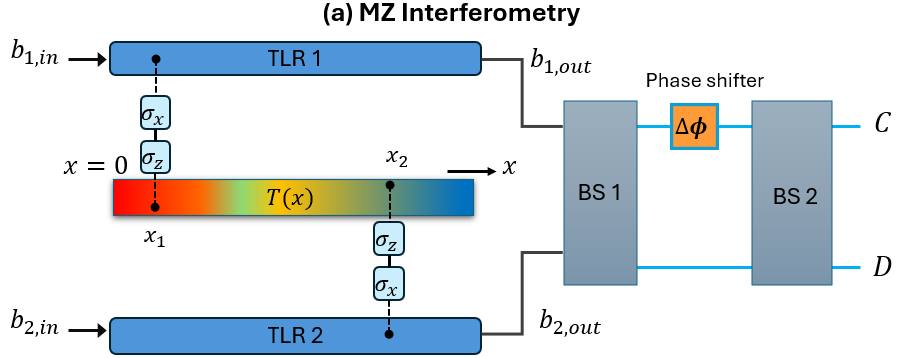}
    	\includegraphics[scale=0.35]{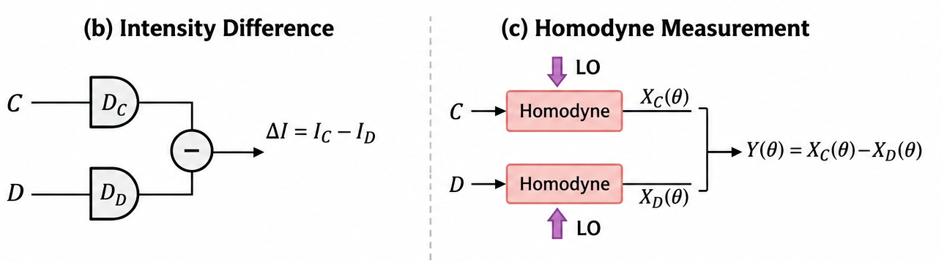}
	\caption{ Circuit diagram of the interferometric temperature-sensing setup. \textbf{(a)} Two transmission line resonators (TLR1 and TLR2), each locally coupled to a probe, sample the spatially varying temperature profile $T(x)$ at positions $x_1$ and $x_2$, encoding temperatures $T_1$ and $T_2$ into the output fields $b_{1,\mathrm{out}}$ and $b_{2,\mathrm{out}}$, originating from inputs $\hat{b}_{1,\mathrm{in}}$ and $\hat{b}_{2,\mathrm{in}}$. The output fields are combined at a first hybrid coupler (BS1), converting them into superpositions suitable for interferometric processing. A relative phase shift $\Delta\phi$ is applied between the arms, followed by recombination at a second hybrid coupler (BS2), which implements the Mach--Zehnder interferometer transformation and maps phase information onto the output modes $C$ and $D$. \textbf{(b)} The MZI output modes $\hat{C}$ and $\hat{D}$ are measured via intensity difference detection, yielding the photocurrent  $\Delta I = I_C-I_D$.  \textbf{(c)} Alternatively, the output modes are also read out via homodyne detection, where the output modes $\hat{C}$ and $\hat{D}$ are each mixed with a local oscillator at independent beamsplitters (not shown here) in a balanced homodyne detection scheme, extracting the quadrature observables $\hat{X}_C$ and $\hat{X}_D$ and yielding the combined signal $Y_A = \hat{X}_C - \hat{X}_D$.}
	\label{fig:MZI}
\end{figure}
%------------------------------------------------------------------%
%***********************************************************%
\section{Experimental implementation}
\label{MZI_readout}
%***********************************************************%

The asymmetric $\sigma_z\sigma_x$ interaction of Eq.~\eqref{model1} is not tied to a specific platform: it can be realized in optomechanical systems~\cite{PhysRevA.51.2537,facchi2021spectral}, superconducting circuits~\cite{PhysRevA.69.062320,PhysRevA.75.032329}, cavity QED~\cite{PhysRevA.82.042327,Scala_2020,PhysRevA.104.013722,lonigro2022self}, single three-level atoms in a cavity~\cite{Knight_1986,Phoenix:90}, and nitrogen--vacancy centers in diamond~\cite{y14n-44h3,Rovny2022,Rovny2025}. In each case, the probe's temperature-dependent steady state couples to a surrounding electromagnetic mode, transducing thermal information from the TLS into a propagating bosonic field. We exploit this by embedding the probes in a Mach--Zehnder interferometer (MZI) readout, which we detail below for a superconducting-circuit realization.
%------------------------------------------------------------%
\begin{figure*}[t!]
	\centering
	\subfloat[]{
		\includegraphics[scale=0.33]{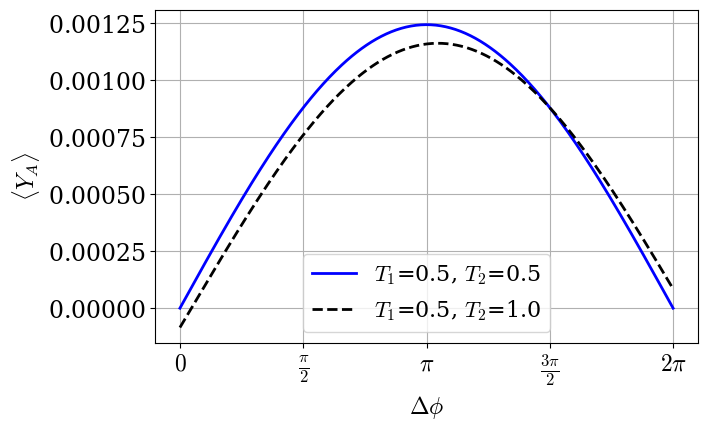}}
	\subfloat[]{
		\includegraphics[scale=0.33]{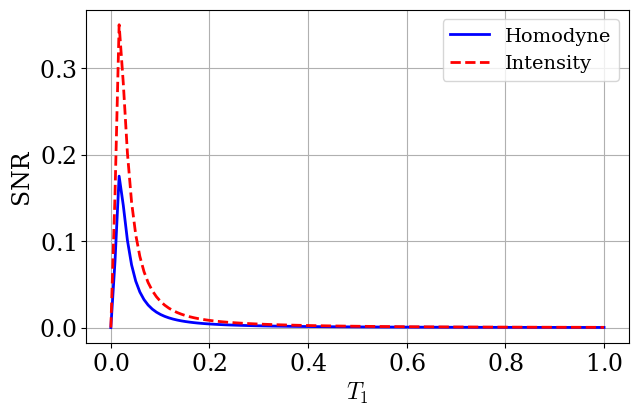}}
	\subfloat[]{
		\includegraphics[scale=0.33]{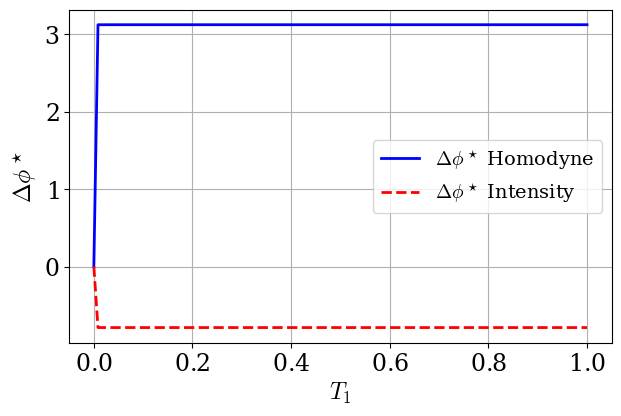}}
	\caption{(a) Phase-dependent homodyne signal $\langle Y_A(\theta)\rangle$ as a function of phase difference $\Delta\phi$. The parameters are set to $\alpha_1=\alpha_2=0.001$, and $\theta=\pi/4$ and are defined in Appendix~\ref{Appendix_A}. (b) Optimized SNR for homodyne (solid blue) and intensity (dashed orange) detection as a function of $T_1$ when $T_2=0.1$. (c) Optimal phase $\Delta\phi^{*}$ that maximizes the SNR for homodyne and intensity detection as a function of $T_1$ when $T_2=0.1$. We fix $\alpha_1=1$, $\alpha_2=1$ in (b) and (c). The rest of the parameters are fixed to $\omega_p=1$, $\omega_a=0.04$, $g=0.04$, $\gamma=0.04$, and $\tau_f=1$.}
	\label{fig:MZI_Results}
\end{figure*}
%------------------------------------------------------------%
As sketched in Fig.~\ref{fig:MZI}, two probe modules sit at positions $x_1$ and $x_2$ along the sample, sensing local temperatures $T_1=T(x_1)$ and $T_2=T(x_2)$. Each probe is coupled to a transmission-line resonator (TLR), which is itself coupled to an open microwave line. The resonator acts as a fast intermediary: when it responds much faster than the probe (the standard bad-cavity limit), it can be eliminated from the dynamics, leaving the probe effectively coupled directly to the propagating line at an effective rate $\gamma$ (See Eq.~\eqref{eq:app_io_effective}). The result is the familiar waveguide-QED picture in which the outgoing field is simply the incoming field plus whatever the probe itself has radiated — precisely what allows the temperature imprinted on the probe's steady state to travel out as a measurable microwave signal.

Since any real detector integrates the field over a finite window rather than sampling it instantaneously, we work with time-filtered field and probe operators that obey this same input--output relation (See Eq.~\eqref{eq:input-output} of Appendix\ref{Appendix_A}); this simply formalizes what ``the measured signal'' means, through an effective detection time $\tau_f$, without changing the physics.

The two arms' outgoing fields are combined on the MZI itself: a pair of $50{:}50$ hybrid couplers with a tunable phase shifter fixing the relative phase $\Delta\phi$ between the arms, producing two output ports $C$ and $D$ (See Eq.~\eqref{C+D} of Appendix\ref{Appendix_A}). Because both ports mix the two arms together, their signals depend on the \emph{difference} between what happened in each arm — that is, on the contrast between $T_1$ and $T_2$ — rather than on either temperature separately. This is what turns the interferometer into a differential thermometer: fixing one probe as a reference and scanning the position of the other measures local temperature distinguishability directly, which is precisely the quantity behind the spatial-resolution analysis of Sec.~III.

The interferometer output can be read out in two complementary ways. \emph{Direct intensity detection} measures the photon flux at the two ports and forms the difference $\Delta I := I_C - I_D$ (Supplemental Material, Sec.~S4). \emph{Homodyne detection} instead mixes each output with a strong local-oscillator tone of phase $\theta$, giving direct access to a chosen field quadrature and to the differential signal $Y_A(\theta)$ (Eq.~\ref{eq:SNR_readout} of Appendix\ref{Appendix_A}). The distinction matters physically: intensity detection is quadratic in the field and is therefore mainly sensitive to \emph{populations}, whereas homodyne detection is linear in the field and gives direct access to the temperature-dependent \emph{probe coherence}. Since it is precisely this coherence that drives the low-temperature quantum enhancement identified earlier through the QFI, homodyne detection is the readout that couples most directly to the thermometric resource being exploited here — although, as we show below, it is not always the more sensitive one in absolute terms.

To compare the two readouts on equal footing, we ask how well each resolves a small change in $T_1$ at fixed reference temperature $T_2$. What matters is not a signal's mean value alone but how strongly it responds to $T_1$ relative to its own fluctuations; we quantify this with the signal-to-noise ratio (SNR) of each observable (Eq.~\eqref{eq:SNR_readout} of Appendix\ref{Appendix_A}). A larger SNR means a smaller resolvable temperature change, and after $\nu$ repeated measurements the achievable resolution improves as $1/\sqrt{\nu}$. For each readout, at every $T_1$ we optimize the interferometer phase $\Delta\phi$ to maximize its SNR.

Figure~\ref{fig:MZI_Results}(a) shows the mean homodyne signal versus interferometric phase for two representative temperature pairs: an approximately sinusoidal curve, as expected for a Mach--Zehnder geometry, whose visibility and offset can be tuned via the local-oscillator phase $\theta$ and the input field amplitudes. Changing the temperatures leaves the curve's overall shape essentially unchanged but shifts its slope near a given operating point — and it is this slope, not the raw signal, that ultimately sets the achievable sensitivity.

The optimized sensitivities in Fig.~\ref{fig:MZI_Results}(b) show a pronounced low-temperature maximum for both readouts, tracking the physics identified earlier in the paper: at very low $T_1$ the probe is essentially frozen and barely responds to further cooling, so sensitivity vanishes; as $T_1$ rises, the auxiliary energy scale becomes thermally active, the probe's response grows rapidly, and sensitivity peaks — in the same low-temperature window where the QFI showed its coherence-assisted enhancement. At higher temperatures, the probe's populations and coherences vary more slowly while measurement noise stays finite, so sensitivity falls again. For the parameters shown, intensity detection reaches a somewhat higher peak sensitivity than homodyne detection, but this does not make homodyne the weaker choice: the two readouts probe the field differently, with intensity mixing direct power and interference contributions while homodyne accesses a field quadrature directly and so couples more cleanly to the coherence-carrying part of the signal. The two are complementary rather than competing.

Figure~\ref{fig:MZI_Results}(c) shows the optimal interferometric phase for each observable. After a brief low-temperature transient, both settle to nearly constant values, $\Delta\phi^\star\simeq\pi$ for homodyne and $\Delta\phi^\star\simeq-\pi/4$ for intensity detection — an experimentally convenient result, since the interferometer can then run near a fixed operating point across most of the relevant temperature range without continuous re-tuning. The homodyne optimum has a simple interpretation: $\Delta\phi=\pi$ is the balanced configuration in which the differential quadrature is maximally sensitive to the relative field emitted by the two probes; the intensity optimum differs because its signal mixes direct and interference contributions in a different combination.

Taken together, these results show how the microwave interferometer converts local temperature information into an accessible signal using standard circuit-QED hardware: transmission-line resonators, microwave hybrid couplers, a tunable phase shifter, and homodyne or intensity detection chains, all routinely available in current experiments. Keeping one probe at a reference position while scanning the other yields the local temperature $T(x)$ through a calibrated response, with our analysis identifying which readout and operating phase deliver the best sensitivity in a given temperature range.
%********************************************************************%

%********************************************************************%

%*****************************************************************************%
\section{Concluding remarks} 
\label{conclusion}

In this work, we examined coherence-enhanced quantum thermometry and its impact on the spatial resolution of one-dimensional temperature profiles. Building on a probe--auxiliary TLS architecture~\cite{PhysRevResearch.5.043184}, we showed that the low-temperature enhancement of the QFI originates from temperature-dependent coherences in the probe's local basis: these coherences contribute an additional term to the QFI beyond the population-only contribution, sharpening the distinguishability of spatial temperature variations.

To benchmark the protocol and assess its general applicability, we considered three representative temperature profiles. For the linear-gradient case, we showed explicitly how this coherence-enabled enhancement translates into a smaller minimum resolvable spatial distance: relative to the corresponding incoherent (locally dephased) probe, in which coherences are suppressed, the coherence-assisted protocol reduced this minimum resolvable distance by orders of magnitude in the regimes considered. These results establish coherence as a key resource for enhanced spatial resolution in quantum thermometry. We further examined how performance scales with the number of auxiliary TLSs. The interaction renormalizes the auxiliary-sector gap, so the array tunes its optimal working temperature and extends the useful window below the bare auxiliary frequency. Optimizing the working temperature at each N yields a finite-size power
law that saturates near $N \sim 50$; under a fixed total coupling resource the advantage disappears. The gain is therefore one of gap engineering rather than metrological scaling.

Finally, we proposed an interferometric readout scheme based on an MZI for extracting spatial temperature information from two probe channels: by injecting the temperature-dependent output fields of the probe--TLR modules into the interferometer, temperature differences between two spatial locations are mapped onto a measurable interference signal at its output ports. Comparing intensity-difference and homodyne detection through their signal-to-noise ratio, we found that both interferometric schemes convert local thermal responses into experimentally accessible observables with robustness and precision.

Beyond the single-quadrature scheme considered here, double homodyne detection could access both quadratures simultaneously, enabling a more complete reconstruction of the output field. Moreover, while the present analysis is restricted to one-dimensional temperature profiles, extending it to scanning geometries, imaging configurations, and two-dimensional thermal maps points toward a broader goal: quantum-assisted high-resolution thermal metrology.

\section*{Acknowledgments}
This work is supported by the Scientific and Technological Research Council (TÜBİTAK) of T\"urkiye under Project Grant No. 123F150, the European Union's Horizon Europe research and innovation programme under the Project ``Quantum Secure Networks Partnership'' (QSNP, Grant Agreement No.~101114043), by INFN through the project ``QUANTUM'' and by Agencia Estatal de Investigaci\'on (Grants No~PID2021-127781NB-I00 and PID2025-172975NB-I00).  G. S. and L. L. S.-S. thank Ko\c{c} University for its kind hospitality.
%**********************************************************%
\appendix
%**********************************************************%
\section{Interferometric readout scheme}\label{Appendix_A}
We provide the derivations underlying the interferometric readout scheme presented in the main text. We start with the microscopic model describing each sensing arm. The corresponding full Hamiltonian can be written as 
\begin{equation}
	H_{\mathrm{arm}}^{i}=H_{(i)}+H_{\mathrm{TLR}}^{(i)}+H_{\mathrm{line}}^{(i)}+H_{p-r}^{(i)}+H_{r-\ell}^{(i)} \, ,
\end{equation}
with resonator, line, probe--resonator, and resonator--line terms given as~\cite{PhysRevA.69.062320,Wallraff2004,PhysRevA.104.013722}
\begin{align}
	H_{\mathrm{TLR}}^{(i)}  & = \hbar\omega_r \, a_i^\dagger a_i \, , \\
	H_{\mathrm{line}}^{(i)} & = \int  d\omega\ \hbar \omega\, b^\dagger(\omega)b(\omega) \, ,\\
	H_{p-r}^{(i)} & = \hbar g_r \!(a_i^\dagger\sigma_-^{(i)}+a_i\sigma_+^{(i)} ) \, ,\\
	H_{r-\ell}^{(i)} & =\int d\omega\,\hbar \kappa(\omega)\! [a_i^\dagger b_i(\omega)+a_ib_i(\omega)^\dagger ] \, .
\end{align}
Using the standard input-output relation, this microscopic  model allows us to write the output field in each arm  as~\cite{PhysRevA.31.3761,RevModPhys.82.1155}
\begin{equation}\label{eq:input-output}
	b_{\mathrm{out}}(t)=b_{\mathrm{in}}(t)+\sqrt{\kappa}\,a(t) \, ,
\end{equation}
i.e., the output field is the incident field plus the field radiated by the resonator. When the resonator serves as a fast readout element, its dynamics can be adiabatically eliminated: near resonance and in the bad-cavity limit $\kappa\gg g_r$, we get
\begin{equation}
	a(t) \simeq -\frac{2}{\sqrt{\kappa}}\,b_{\mathrm{in}}(t)-i\frac{2g_r}{\kappa}\,\sigma_-(t) \, ,
\end{equation}
up to an overall phase convention. Substituting back into the input--output relation and absorbing irrelevant phases gives the effective relation used throughout the main text,
\begin{equation}
	\label{eq:app_io_effective}
	b_{\mathrm{out}}(t)=b_{\mathrm{in}}(t)+\sqrt{\gamma}\,\sigma_-(t) \, , \qquad \gamma\simeq\frac{4g_r^2}{\kappa} \, ,
\end{equation}
for both arms. Therefore,  $b^{(1)}{\mathrm{out}}$ and $b^{(3)}{\mathrm{out}}$  each carry information about $T_1$ and $T_2$ through the corresponding probe steady state.

Since detection selects a finite temporal mode rather than an instantaneous field, we introduce normalized filtered wavepacket operators,
\begin{equation}
	B^{(i)}_{\mathrm{out/in}}=\int dt\,f(t)\,b^{(i)}_{\mathrm{out/in}}(t) \, , \qquad \int dt\,|f(t)|^2=1 \, ,
\end{equation}
and the corresponding filtered source operator
\begin{equation}\label{Si}
	S^{(i)} :=\int dt\,f(t)\,\sigma_-^{(i)}(t) \, ,
\end{equation}
so that
\begin{equation}\label{eq:app_io_filtered}
	B^{(i)}_{\mathrm{out}}=B^{(i)}_{\mathrm{in}}+\sqrt{\gamma}\,S^{(i)} \, .
\end{equation}
The filtered probe operators $S_i$ are what ultimately feed the interferometer.

The outgoing modes $B^{(1)}_{\mathrm{out}}$ and $B^{(2)}_{\mathrm{out}} $ enter a microwave Mach-Zehnder interferometer (MZI), with $50{:}50$ hybrid couplers as beam splitters and a tunable phase shifter setting the arm phase difference. With
\begin{equation}
	U_{\mathrm{BS}}=\frac{1}{\sqrt2}\begin{pmatrix}1&i\\i&1\end{pmatrix} , \qquad
	U_\phi= \begin{pmatrix} e^{i \phi_1}&0\\0&e^{i\phi_2}\end{pmatrix}  ,
\end{equation}
the full transformation $U_{\mathrm{MZI}}=U_{\mathrm{BS}}U_\phi U_{\mathrm{BS}}$ reads
\begin{equation}
	U_{\mathrm{MZI}} = e^{i\bar\phi}
	\begin{pmatrix}
		i\sin(\Delta\phi/2) & i\cos(\Delta\phi/2) \\
		i\cos(\Delta\phi/2) & -i\sin(\Delta\phi/2)
	\end{pmatrix} \, ,
\end{equation}
with $\Delta \phi = \phi_2 - \phi_1$. This gives output modes
\begin{equation}\label{C+D}
	\begin{pmatrix}C\\D\end{pmatrix} = U_{\mathrm{MZI}}\begin{pmatrix}B^{(1)}_{\mathrm{out}}\\B^{(2)}_{\mathrm{out}}\end{pmatrix} \, .
\end{equation}
and, therefore, output intensities $I_C= \langle C^\dagger C\rangle$ and $I_D=\langle D^\dagger D\rangle$ are
\begin{equation}
\begin{aligned}
I_C &= \sin^2\!\left(\frac{\Delta\phi}{2}\right)n_1 +\cos^2\!\left(\frac{\Delta\phi}{2}\right)n_2 +\frac{1}{2}\sin(\Delta\phi)\,X,\\[1mm]
I_D &= \cos^2\!\left(\frac{\Delta\phi}{2}\right)n_1 +\sin^2\!\left(\frac{\Delta\phi}{2}\right)n_2 -\frac{1}{2}\sin(\Delta\phi)\,X,
\end{aligned}
\end{equation}
where $n_i=\langle B^{(i) \dagger} B^{(i)} \rangle$ and $X=2\,\mathrm{Re}\,\langle  B^{(1) \dagger}   B^{(2)} \rangle$ for output modes. The terms proportional to $n_1$ and $n_2$ are the direct contributions from the two arms, while the term proportional to $X$ is the interference contribution. The natural intensity-based observable used in the main text is the difference
\begin{equation}
	\Delta I := I_C-I_D = -\cos(\Delta\phi)\,(n_1-n_2)+\sin(\Delta\phi)\,X \, .
\end{equation}
Through Eq.~\eqref{C+D}, $n_i$ and $X$ contain probe correlators: once $T_1$ and $T_2$ fix the probe steady states, they fix the outgoing fields and hence the interferometric signal.

We consider homodyne detection, as it is perhaps the standard technique in continuous variables. It mixes the output with a strong local-oscillator tone at the same frequency, whose phase $\theta$ selects the measured quadrature,
\begin{equation}\label{X_C+X_D}
	X_{C,D}(\theta)=\frac{1}{2}(O\,e^{-i\theta}+O^\dagger e^{i\theta} ) \, , \qquad O=C,D \, .
\end{equation}
We consider the differential homodyne signal $Y_A(\theta):=X_C(\theta)-X_D(\theta)$, with mean value
\begin{equation}\label{homodyne-main}
	\langle Y_A(\theta)\rangle = \frac{1}{2}\,\mathrm{Re}\!\left[e^{-i\theta}\langle C-D\rangle\right] \, ,
\end{equation}
which makes the phase sensitivity explicit: varying $\theta$ selects which component of the complex amplitude $\langle C-D\rangle$ is measured. From Eq.~\eqref{C+D},
\begin{equation}
	\langle C-D\rangle = \kappa_1(\Delta\phi) (\alpha_1+\sqrt{\gamma} \,\langle S^{(1)} \rangle ) + \kappa_2(\Delta\phi)\ \alpha_2+\sqrt\gamma\,\langle S^{(2)}\rangle ) \, ,
\end{equation}
with
\begin{equation}
\begin{aligned}
	\kappa_1(\Delta\phi) &= i\!\left[\sin\!\left(\tfrac{\Delta\phi}{2}\right)-\cos\!\left(\tfrac{\Delta\phi}{2}\right)\right] \, ,  \\
	\kappa_2(\Delta\phi) &= i\!\left[\cos\!\left(\tfrac{\Delta\phi}{2}\right)+\sin\!\left(\tfrac{\Delta\phi}{2}\right)\right] \, ,
	\end{aligned}
\end{equation}
and $\alpha_i:=\langle B^{(i)}_{\mathrm{in}}\rangle$ is the coherent input amplitude in arm $i$. The signal depends linearly on $\langle S_i\rangle$ — and so directly on the temperature-dependent probe coherence — which is why homodyne readout accesses the coherence-carrying part of the signal directly.

To complete the link with thermometry, we take the total input state to be
\begin{equation}
\varrho_0=\varrho_{\mathrm{field}}\otimes \varrho_1(T_1)\otimes \varrho_2(T_2) \, ,
\end{equation}
where $\varrho_{\mathrm{field}}$ describes the incoming microwave modes and $\varrho_i(T_i)$ is the steady state of probe $i$ at the local temperature $T_i=T(x_i)$. The relevant temperature-dependent probe observables are 
\begin{equation}
\mu_i(T_i):=\langle \sigma_-^{(i)}\rangle_{T_i}=c_i(T_i) \, ,
\qquad
p_i(T_i):=\langle \sigma_+^{(i)}\sigma_-^{(i)}\rangle_{T_i} \, .
\end{equation}
In the steady-state narrowband limit,
\begin{equation}
	\langle S^{(i)} \rangle \simeq \mu_i(T_i)\,\tau_f \, , \qquad
	\langle S^{(i) \dagger} S_i\rangle \simeq p_i(T_i)\,\tau_f \, , 
\end{equation}
with $\tau_f  :=\int dt\,f(t)$ the effective detection window. Both the intensity and homodyne readouts are therefore ultimately functions of $T_1=T(x_1)$ and $T_2=T(x_2)$: for fixed positions, $I_C$, $I_D$, $\Delta I$, and $Y_A(\theta)$ measure the difference between the two local thermal responses.

For an observable $O\in\{\Delta I, Y_A\}$, standard error propagation gives the smallest detectable temperature change as
\begin{equation}\label{eq:dTmin_readout}
	\delta T_{1,\min}^{(O)} = \frac{\sqrt{\mathrm{Var}(O)}}{\left|\partial_{T_1}\langle O\rangle\right|} \, ,
\end{equation}
noise over response — smaller is better. Equivalently, we define the signal-to-noise ratio
\begin{equation}\label{eq:SNR_readout}
	\mathrm{SNR}_O(T_1) := \frac{\left|\partial_{T_1}\langle O\rangle\right|}{\sqrt{\mathrm{Var}(O)}} \, ,
\end{equation}
so that larger $\mathrm{SNR}_O$ means finer resolution; after $\nu$ repetitions, $\delta T_{1,\min}^{(O,\nu)} = 1/(\sqrt\nu\,\mathrm{SNR}_O(T_1))$. For homodyne detection we fix the local-oscillator phase to maximize the quadrature's temperature response, then optimize the remaining interferometric phase,
\begin{equation}
	\Delta\phi_O^\star(T_1) = \arg\max_{\Delta\phi\in[0,2\pi]}\mathrm{SNR}_O(T_1) \, ,
\end{equation}
with $O=\Delta I$ for intensity readout and $O=Y_A$ for homodyne readout. These are the quantities plotted in Fig.~{9} of the main text.
\bibliography{SR_L}

\end{document}